\documentclass[a4paper,12pt,english]{article}
\usepackage{amssymb}
\usepackage{amsfonts}
\usepackage{graphicx}
\usepackage{amsmath}
\usepackage{color}

\newcommand{\be}{\begin{equation}}
\newcommand{\ee}{\end{equation}}
\begin{document}


\begin{titlepage}
\begin{center}

\noindent{{\LARGE{Screening Stringy Horizons}}}

\smallskip
\smallskip

\smallskip
\smallskip
\smallskip
\smallskip
\noindent{\large{Gaston Giribet$^{1,2,3}$, Arash Ranjbar$^{1,4}$}}

\smallskip
\smallskip

\end{center}

\smallskip
\smallskip
\centerline{$^1$ Universit\'{e} Libre de Bruxelles and International Solvay Institutes}
\centerline{{\it ULB-Campus Plaine CPO231, B-1050 Brussels, Belgium.}}

\smallskip
\smallskip
\centerline{$^2$ Departamento de F\'{\i}sica, Universidad de Buenos Aires and IFIBA-CONICET}
\centerline{{\it Ciudad Universitaria, Pabell\'on 1, 1428, Buenos Aires, Argentina.}}

\smallskip
\smallskip
\centerline{$^3$ Instituto de F\'{\i}sica, Pontificia Universidad Cat\'{o}lica de
Valpara\'{\i}so}
\centerline{{\it Casilla 4059, Valpara\'{\i}so, Chile.}}

\smallskip
\smallskip
\centerline{$^4$ Centro de Estudios Cient\'{\i}ficos (CECs)}
\centerline{{\it Arturo Prat 514, Valdivia, Chile.}}

\bigskip

\bigskip

\bigskip

\bigskip

\begin{abstract}

It has been argued recently that string theory effects qualitatively modify the effective black hole geometry
experienced by modes with radial momentum of order $1/\sqrt{\alpha '}$. At tree level, these ${\alpha '}$-effects
can be explicitly worked out in two-dimensional string theory, and have a natural explanation in the T-dual
description as coming from the integration of the zero-mode of the linear dilaton, what yields a contribution
that affects the scattering phase-shift in a peculiar manner. It has also been argued that the phase-shift
modification has its origin in a region of the moduli space that does not belong to the exterior black hole
geometry, leading to the conclusion that at high energy the physics of the problem is better described by the dual model. Here, we elaborate on this argument. We consider the contribution of worldsheet instantons in the 2D Euclidean black hole $\sigma $-model and study its influence on the phase-shift at high energy.

\end{abstract}

\end{titlepage}


\section{Introduction}

In the recent paper \cite{este}, it has been argued that string theory
effects qualitatively modify the effective black hole geometry experienced
by modes with radial momentum of order $1/\sqrt{\alpha ^{\prime }}$. At tree
level, these $\alpha ^{\prime }$-effects can be explicitly worked out in
two-dimensional string theory, where the black hole background admits an
exact worldsheet description in terms of the gauged $SL(2,\mathbb{R})/U(1)$
Wess-Zumino-Witten (WZW) model. In addition, this model is known to have a
dual description (in a sense similar to T-duality) that involves a two-dimensional flat
tachyonic linear dilaton background, known as
Fateev-Zamolodchikov-Zamolodchikov (FZZ) dual \cite{FZZ}. In the FZZ dual
model, the $\alpha ^{\prime }$ effects studied in \cite{este} have a natural
explanation as coming from the integration over the zero-mode of the linear
dilaton, what yields a contribution that affects the scattering phase-shift
in a peculiar manner. Such contributions and, consequently, the phase-shift
modification they produce seem to come from a region of the moduli
space that does not belong to the exterior black hole geometry. This led the
authors of \cite{este} to conclude that the high-energy physics is better
described by the FZZ\ dual model, rather than by the two-dimensional (2D) black hole $\sigma $%
-model. Here, we elaborate on this argument. We consider worldsheet instanton contributions and study their influence to the phase-shift at high energy. The 2D black hole $\sigma $-model contains an operator that controls the worldsheet instanton contributions. This is given by a {\it second} screening operator in the worldsheet CFT, and the integration over the zero-mode of the linear dilaton, once such screening operator is considered, yields an $\alpha '$-dependent function of the string states momenta. Therefore, it is natural to ask to what extent such $\alpha '$-dependent function accounts for the high-energy phase-shift discussed in \cite{este}. We will study this in Section 5. Before, in Section 2, we review the 2D Euclidean black hole $\sigma $-model. In Section 3, we discuss string scattering amplitudes on the black hole background and the high-energy modification of the phase-shift. In Section 4, we review how such modification is gathered in the FZZ dual theory. In Section 5, we first show how the second screening is associated to worldsheet instantons, and then study their contribution in relation to the stringy phase-shift.

\section{String 2D black hole}

String theory admits an interesting exact solution that describes strings
propagating on a two-dimensional black hole background \cite{Mandal, W}. The
metric of the classical Euclidean black hole geometry is given by%
\begin{equation}
ds^{2}=L^{2}\left( dr^{2}+\tanh ^{2}(r)\ d\theta ^{2}\right) ,  \label{G}
\end{equation}%
where $\theta \in \lbrack 0,2\pi )$ and $r\in \mathbb{R}_{\eqslantgtr 0}$.
On the other hand, the dilaton configuration needed to support this
background is given by%
\begin{equation}
\Phi (r)=\Phi _{0}-\log \left( \cosh (r)\right) ,  \label{D}
\end{equation}%
which tends linearly to infinity when $r$ is large, namely far from the
horizon. The metric above describes a semi-infinite cigar-like geometry of $%
\mathbb{R}^{2}$ topology. Compact coordinate $\theta $ represents the
compactified Euclidean time. The horizon of the Lorentzian black hole gets
mapped to the tip of the cigar, which is located at $r=0$. In the large $r$
region, the Euclidean metric approaches a cylinder of radius $L$.

The non-linear $\sigma $-model that describes 2D string
theory propagatig on the background (\ref{G})-(\ref{D}) is given by the
gauged Wess-Zumino-Witten (WZW)\ model for the coset $H^+_3/U(1)$, with $H^+_3=SL(2,\mathbb{C})/SU(2)$. Its Lorentzian version corresponds to the $SL(2,\mathbb{R})/U(1)$ coset
\cite{W}. Let us first discuss the $\sigma $-model on $H^+_3$,
which in a convenient coordinate system takes the form%
\begin{equation}
S_{M}=\frac{L^{2}}{2\pi }\int d^{2}z\left( \partial \phi \bar{\partial }%
\phi +\beta \bar{\partial }\gamma +\bar{\beta }\partial \bar{%
\gamma }-\frac{R\phi }{2\sqrt{2(k-2)}}-2\pi M\ \beta \bar{\beta }e^{-%
\sqrt{\frac{2}{k-2}}\phi }\right)   \label{sigma}
\end{equation}%
which involves a scalar field $\phi $ and a $\beta $-$\gamma $ commuting
ghost system. At large $r$, one can identify $\phi \sim \sqrt{k}r$, and this
corresponds to a linear dilaton background. The constant $M$ appearing in (%
\ref{sigma})\ can be shown to be related to the two-dimensional black hole
mass. Its precise value is not relevant for perturbative physics, as it can
be set to 1 by simply rescaling $\phi $. The relevant parameter is actually 
\begin{equation*}
k=\frac{L^{2}}{\alpha ^{\prime }},
\end{equation*}%
which is the Reynolds number that controls the physics of the problem:\ It
measures the typical strings length, $l_{s}=\sqrt{\alpha ^{\prime }}$,
relative to the curvature radius of the geometry, $L$. The classical limit
thus corresponds to $k$ going to infinity.

Ghost fields $\beta $ and $\bar{\beta }$ in (\ref{sigma}) are not
dynamical, and then they can be integrated out. However, it is convenient to
keep these fields in the action for practical purposes. One of the reasons
for doing this is that it allows for a free field representation of the $sl(2,%
\mathbb{R})_{k} \oplus sl(2,\mathbb{R})_{k}$ Kac-Moody symmetry that action (%
\ref{sigma}) exhibits. This free field realization is given by the so-called Wakimoto
representation of the $sl(2,\mathbb{R})_{k}$ current algebra, which follows from defining the local currents \cite{Wakimoto}
\begin{eqnarray}
J^{+}(z) &=&\beta (z),  \label{J1} \\
J^{3}(z) &=&-\beta (z)\gamma (z)-\sqrt{\frac{k-2}{2}}\partial \phi (z),
\label{J2} \\
J^{-}(z) &=&\beta (z)\gamma ^{2}(z)+\sqrt{2k-4}\gamma (z)\partial \phi
(z)+k\partial \gamma (z),  \label{J3}
\end{eqnarray}%
together with their complex conjugate counterpart. Considering the free-field
propagators both for the scalar field $\phi $ and for the $\beta $-$\gamma $
system, the operator product expansion (OPE)\ of currents (\ref{J1})-(\ref%
{J3}) realizes the $sl(2,\mathbb{R})_{k}$ affine Kac-Moody algebra. It can
be verified that the interaction term in (\ref{sigma}) has a regular OPE\
with the currents, so it preserves the full $sl(2,\mathbb{R})_{k}$ symmetry.

Through the Sugawara construction, the currents (\ref{J1})-(\ref{J3}) yield
the stress-tensor%
\begin{equation}
T_{SL(2,\mathbb{R})}=\beta (z)\partial \gamma (z)-\frac{1}{2}(\partial \phi
(z))^{2}-\frac{1}{\sqrt{2(k-2)}}\partial ^{2}\phi (z),  \label{pseudoT}
\end{equation}%
together with its anti-holomorphic counterpart.

The coset $SL(2,\mathbb{R)}/U(1)$ construction can be accomplished by
supplementing the $SL(2,\mathbb{R})$ model by adding an extra scalar field $%
X(z)=X_{L}(z)+X_{R}(\bar{z})$ and a fermionic $B$-$C$ ghost system \cite{BK,DVV,BB}.
This amounts to improve the stress-tensor (\ref{pseudoT}) with an extra
piece, namely\footnote{Hereafter we omit the diffeomoerphism $b$-$c$ ghost system
contributions.} 
\begin{equation}
T_{SL(2,\mathbb{R})/U(1)}=T_{SL(2,\mathbb{R})}-B(z)\partial C(z)-\frac{1}{2}%
(\partial X(z))^{2},  \label{T}
\end{equation}%
and analogously for the anti-holomorphic counterpart. This yields the
central charge%
\begin{equation*}
c=\frac{2k+2}{k-2},
\end{equation*}%
which consistently tends to $2$ in the large $k$ limit.

The BRST charge associated to the $U(1)$ of the coset model is%
\begin{equation}
Q_{\text{BRST}}^{U(1)}=\int dz\ C(z)\left( J^{3}(z)-i\sqrt{k/2}\partial
X(z)\right) .  \label{BRST}
\end{equation}%
This means that the vertex operators creating physical states of the theory
have to have regular OPE with the current $J^{3}-i\sqrt{k/2}\partial X$.
Such operators are%
\begin{equation}
V_{j,m,\bar{m}}(z,\bar{z})=\gamma ^{j-m}(z)\bar{\gamma }^{j-%
\bar{m}}(\bar{z})e^{\sqrt{\frac{2}{k-2}}j\phi (z,\bar{z})}e^{i%
\sqrt{\frac{2}{k}}(mX_{L}(z)+\bar{m}X_{R}(\bar{z}))},  \label{V}
\end{equation}%
where $j$, $m$ and $\bar{m}$ are isospin variables that label the $SL(2,%
\mathbb{R)\times }SL(2,\mathbb{R)}$ representations. These variables
represent the momenta associated to the radial and Euclidean time
coordinates, and the winding number along the latter. More precisely, we
have the radial momentum%
\begin{equation}
p_{\phi }\equiv -i\frac{2j+1}{\sqrt{k-2}}
\end{equation}%
and the right- and left-moving momenta%
\begin{eqnarray}
p_{L} &\equiv &\frac{2}{\sqrt{k}}m=\left( \omega \sqrt{k}+p_{\theta
}\right) , \\
p_{R} &\equiv &\frac{2}{\sqrt{k}}\bar{m}=\left( \omega \sqrt{k}%
-p_{\theta }\right) ,
\end{eqnarray}%
where $p_{\theta }$ and $\omega $ represent the momentum and the winding
number along the $\theta $-direction, respectively. That is to say that the
vertices (\ref{V}) go like $V\sim e^{\frac{i}{\sqrt{2}}(p_{\phi }-iQ)\phi
}e^{\frac{i}{\sqrt{2}}(p_{L}X_{L}+p_{R}X_{R})}$, with $Q=-1/\sqrt{k-2}$.

States with $p_{\phi }\in \mathbb{R}$ and $m-\bar{m}\in \mathbb{Z}$
correspond to vectors of the continuous series representation of $SL(2,%
\mathbb{R})$. Discrete representations have $j\in \mathbb{R}$, $m-\bar{m%
}\in \mathbb{Z}$ and $m+\bar{m}-j\in \mathbb{Z}$. Here, we will be involved with the former. 

\section{High-energy scattering}

Now, let us discuss string scattering amplitudes on the background (\ref{G})-(\ref{D}). Tree-level string scattering amplitudes are given by the integral of
correlation functions of vertex operators on the sphere topology%
\begin{equation}
A_{m_{1}m_{2}...m_{N}}^{j_{1}j_{2}...j_{N}}=\int
\prod\limits_{i=1}^{N}d^{2}z_{i}\text{Vol}_{SL(2,\mathbb{C)}%
}^{-1}\left\langle \prod\limits_{\ell=1}^{N}:V_{j_{\ell},m_{\ell},\bar{m}%
_{\ell}}(z_{\ell},\bar{z}_{\ell}):\right\rangle _{M}  \label{A1}
\end{equation}%
where Vol$_{SL(2,\mathbb{C)}}^{-1}$ is the volume of the conformal Killing
group. The subscript $M$ in (\ref{A1}) refers to the fact that expectation
value is defined by interacting action (\ref{sigma}). More precisely, we can
write 
\begin{equation}
A_{m_{1}m_{2}...m_{N}}^{j_{1}j_{2}...j_{N}}=\int
\prod\limits_{i=4}^{N}d^{2}z_{i}\int \mathcal{D}^{2}\gamma \ \mathcal{D}%
^{2}\beta \ \mathcal{D}\phi \ \mathcal{D}X\
e^{-S_{M}}\prod\limits_{\ell=1}^{N}V_{j_{\ell},m_{\ell},\bar{m}_{\ell}}(z_{\ell},%
\bar{z}_{\ell}),  \label{A2}
\end{equation}%
where the worldsheet insertions of three vertices fixed as $z_{1}=%
\bar{z}_{1}=0$, $z_{2}=\bar{z}_{2}=0$, and $z_{3}=\bar{z}%
_{3}=\infty $ in order to cancel the stabilized factor Vol$_{SL(2,\mathbb{C)}%
}^{-1}$. The functional measures $\mathcal{D}^{2}\gamma $ and $\mathcal{D}^{2}\beta 
$ stand for both the holomorphic and anti-holomorphic contributions of the
ghost system. It can be shown that the integration on $\beta $ and $%
\bar{\beta }$ is what ultimately induces the one-loop corrections in (\ref%
{sigma}), including in particular the linear dilaton term.

To perform the functional integral (\ref{A2}) it is convenient to
first separate the zero-mode $\phi _{0}$ of the field $\phi $. That is,
consider $\phi (z,\bar{z})=\phi _{0}+\tilde{\phi }(z,\bar{z})$
and then integrate over the dilaton zero-mode $\phi _{0}$ and
fluctuations $\tilde{\phi }(z,\bar{z})$ separately. When doing this, one verifies that the integrand in (\ref{A2}) involves a contribution
\begin{equation}
e^{-\tilde{S}_{0}} e^{\sqrt{\frac{2}{k-2}}%
j_{i}\tilde{\phi }(z_{i},\bar{z}_{i})}\int_{\mathbb{R}}d\phi
_{0}\int_{\mathbb{R}_{+}}d\eta \ \delta \left( \eta -e^{-\phi _{0}\sqrt{%
\frac{2}{k-2}}}\tilde{S}_{\text{I}}\right) e^{-\eta }e^{\phi _{0}\sqrt{%
\frac{2}{k-2}}\left( \sum_{i=1}^{N}j_{i}+\frac{1}{4\pi }\int d^{2}zR\right) }  
\label{pi}
\end{equation}%
which comes from separating the action (\ref{sigma})\ in its free part $%
S_{0}\equiv S_{M=0}$ and the interaction term%
\begin{equation}
S_{\text{I}}=M\int d^{2}z \ W(z,\bar{z})\ , \ \ \ \ \  W=\beta (z)  \bar{\beta }(\bar{z})e^{-\sqrt{\frac{2}{k-2}}\phi (z,\bar{z}) } .  \label{SI}
\end{equation}%

In (\ref{pi}), quantities with tilde, such as $\tilde{S}_{0}$ and $%
\tilde{S}_{\text{I}}$, are defined by replacing the field $\phi (z,%
\bar{z})$ by its fluctuations $\tilde{\phi }(z,\bar{z})$, 
\textit{e.g.} $S_{0}=\tilde{S}_{0}$ and $S_{\text{I}}=e^{-\phi _{0}\sqrt{%
\frac{2}{k-2}}}\tilde{S}_{\text{I}}$. Considering the Gauss-Bonnet theorem, which states that the Euler
characteristic on the sphere is $\frac{1}{2\pi }\int d^{2}z\ \sqrt{g}R=2$, and using
properties of the $\delta $-function, one finds
that, after integrating over $\phi _{0}$, (\ref{pi}) takes the form
\begin{equation}
-\Gamma (-s)\sqrt{\frac{k-2}{2}}e^{-\tilde{S}_0}e^{\sqrt{\frac{2}{k-2}}%
j_{i}\tilde{\phi }(z_{i},\bar{z}_{i})}\left( \tilde{S}_{\text{I}%
}\right) ^{s},  \label{epa}
\end{equation}%
where $s=\sum_{i=1}^{N}j_{i}+1$ and where $\Gamma (-s)=\int_{\mathbb{R}%
_{+}}d\eta \ \eta ^{-1-s}e^{-\eta }$. Finally, absorbing a $k$-dependent
factor in the definition of the path integral, one finds%
\begin{equation}
A_{m_{1}m_{2}...m_{N}}^{j_{1}j_{2}...j_{N}}=\Gamma (-s) \ M^{s}\int
\prod\limits_{i=4}^{N}d^{2}z_{i}\int \prod\limits_{r=1}^{s}d^{2}w_{r}\
F_{m_{1}m_{2}...m_{N}}^{j_{1}j_{2}...j_{N}}(z_{1},...z_{N},w_{1},...w_{s})
\label{aa}
\end{equation}%
with%
\begin{equation}
F_{m_{1}m_{2}...m_{N}}^{j_{1}j_{2}...j_{N}}=\left\langle
\prod\limits_{i=1}^{N}:V_{j_{i},m_{i},\bar{m}_{i}}(z_{i},\bar{z}%
_{i}):\prod\limits_{r=1}^{s}: 
W(w_{r},\bar{w}_r)
:\right\rangle _{M=0}  \label{bb}
\end{equation}%
where now the expectation value (\ref{bb})\ is defined with
respect to the free action $S_{0}$. This permits to perform the Wick
contractions resorting to the free field propagators. The insertion of the $s
$ additional integrated vertices $M\int \beta \bar{\beta }e^{-\sqrt{%
\frac{2}{k-2}}\tilde{\phi }}$ in (\ref{bb}) comes from the factor $%
( \tilde{S}_{\text{I}}) ^{s}$ in (\ref{epa}). Recall we have%
\begin{equation}
s=1+\sum\limits_{i=1}^{N}j_{i}.  \label{s}
\end{equation}

Then, we see that the integration over the dilaton zero-mode is responsible for the prefactor $\Gamma (-s)$ in (\ref{aa}). This prefactor is crucial for our discussion. This produces
poles at configurations $s=1+\sum_{i=1}^{N}j_{i}\in \mathbb{Z}_{\geq 0}$,
whose residues can be thought of as \textit{resonant} correlators. The poles $s\in \mathbb{Z}_{> 0}$ admit a physical interpretation similar to that proposed in Ref. \cite{diFK} for the analogous poles in Liouville field theory. 

Formula (\ref{aa}) can also be interpreted within the context of the Coulomb
gas realization of 2D CFT\ correlation functions, where the insertion of $s$
operators $\tilde{S}_{\text{I}}$ correspond to the inclusion of
screening charges needed to satisfy the charge condition imposed by the
presence of the background charge $Q=-1/\sqrt{k-2}$ at infinity.

The black hole mass parameter $M$ in amplitudes (\ref{aa})\ plays the role
of string coupling constant: Its power depends on the genus of the surface
goes like $M^{s+1-g}$ and, as mentioned above, its absolute value is
determined by the zero-mode of the dilaton, \textit{i.e.} $M\sim e^{\Phi
_{0}}$.

When integrating over the $\beta $-$\gamma $ system, and because $\beta $ is
a 1-differential, the Riemann-Noch theorem, once combined with (\ref{s}),
yields exactly the same conservation law obtained from the integration over the zero-mode of
the field $X$, namely $\sum_{i=1}^{N}(m_{i}+\bar{m}_{i})=0$. The condition on the total winding number $\sum_{i=1}^{N}(m_{i}-\bar{m}_{i})$, on the other hand, is more subtle \cite{GN3, MO3}.

The prefactor $\Gamma (-s)$ in (\ref{aa}) can be alternatively obtained by
virtually integrating over the imaginary part of $\phi _{0}$. This produces
a $\delta $-function that selects a precise amount of operators $\tilde{S}%
_{\text{I}}$ from the series expansion of $e^{-S_{\text{I}}}$. More
precisely, the $\delta $-function selects the $s^{\text{th}}$ term $\frac{%
(-1)}{s!}\left( S_{\text{I}}\right) ^{s}$ of the series, with $%
s=\sum_{i=1}^{N}j_{i}+1$. Then, the infrared divergence $\delta (0)\sim
\Gamma (0)$ can be combined with the multiplicity factor $(-1)/s!$ in order
to produce the factor $\Gamma (-s )$ by recalling the formula $%
\lim_{\varepsilon \rightarrow 0}\Gamma (-s+\varepsilon )/\Gamma (\varepsilon
)=(-1)^{s}/\Gamma (s+1)$ for $s\in \mathbb{Z}_{\geq 0}$.

Correlation functions (\ref{aa})-(\ref{bb}) can be explicitly computed for
the cases $N=2$ and $N=3$. The expression for the two-point function is \cite%
{BB}%
\begin{equation}
A_{m\ -m}^{j\ j}=\left( -\pi M\ \frac{\Gamma \left( \frac{1}{%
k-2}\right) }{\Gamma \left( \frac{k-3}{k-2}\right) }\right) ^{2j+1}\frac{%
\Gamma \left( 1-\frac{2j+1}{k-2}\right) }{\Gamma \left( 1+\frac{2j+1}{k-2}%
\right) }\frac{\Gamma (1+j-m)\Gamma (1+j+\bar{m})\Gamma (-2j)}{\Gamma
(-j-m)\Gamma (-j+\bar{m})\Gamma (2j+1)}.  \label{2pf}
\end{equation}

The phase-shift $\delta $ is defined by the reflection coefficients, given by the two-point function as $e^{i\delta }\equiv A_{m\ -m}^{j\ j}$. Since $M$ can be adjusted to absorb the $k$-dependent functions in the first factor of (\ref{2pf}), the only relevant $k$-dependent piece in the two-point function is given by the prefactor 
\begin{equation}
\frac{\Gamma \left( 1-\frac{2j+1}{k-2}\right)}{\Gamma \left( 1+\frac{2j+1}{k-2}\right)}
=\Gamma \left( 1-i\frac{p_{\phi }%
}{\sqrt{k-2}}\right) \Gamma^{-1} \left( 1+i\frac{p_{\phi }%
}{\sqrt{k-2}}\right).  \label{vt}
\end{equation} 

This prefactor is the one responsible for the phase-shift modification at high-energy discussed in \cite{este}. To see this, following \cite{este}, one may resort to the Stirling approximation, which states that for large $|z|$ one has $ \Gamma (z)\simeq \sqrt{{2\pi }/{z}}\left( {z}/{e}\right) ^{z}\ \left( 1+%
\mathcal{O}(1/z)\right)$. Applying this to (\ref{vt}), one finds that in the regime $p_{\phi } \gg \sqrt{k-2}$, one finds
\begin{equation}
e^{i\delta } \simeq 
- e^{-\frac{2i p_{\phi }}{\sqrt{k-2}}( \log p_{\phi } - 1 -\frac{1}{2}\log (k-2))} e^{\frac{i\pi}{2}}  .
\label{phaseshift}
\end{equation}

This gives the high-energy dependence of the phase-shift \cite{este}.

\section{The dual theory}

The two-dimensional $\sigma $-model on the Euclidean 2D\ black hole geometry
is known to be dual to a two-dimensional conformal field theory known as
sine-Liouville theory. The latter theory is defined by the action
\begin{equation}
S_{\lambda }=\frac{1}{2\pi }\int d^{2}z\left( \partial \varphi \bar{%
\partial }\varphi +\partial X\bar{\partial }X-\frac{R\varphi }{2\sqrt{%
2k-4}}+2\pi \lambda \ e^{-\sqrt{\frac{k-2}{2}}\varphi }\cos \left( \sqrt{k/2}%
\tilde{X}\right) \right)   \label{sL}
\end{equation}%
where $\tilde{X}(z,\bar{z})\equiv X_{L}(z)-X_{R}(\bar{z})$ is
the dual to the bosonic field $X(z,\bar{z})=X_{L}(z)+X_{R}(\bar{z})
$. Direction $X$ is compact, while $\varphi $ takes values on the real line.
That is to say that, unlike the 2D Euclidean black hole $\sigma $-model, the
model defined by (\ref{sL}) has topology $\mathbb{R\times }S^{1}$. Thought
of as a string $\sigma $-model action, sine-Liouville theory (\ref{sL})
represents a flat linear dilaton background in the presence of a non-homogeneous
tachyon condensate.

The duality between (\ref{sigma}) and (\ref{sL}) has been conjectured by
Fateev, Zamolodchikov and Zamolodchikov (FZZ)\ in an unpublished work \cite%
{FZZ}, and it has been reviewed and elaborated by Kazakov, Kostov and Kutasov in Ref. \cite{KKK}. It
represents a kind of T-duality. In fact, the supersymmetric version of the
FZZ\ duality actually corresponds to mirror symmetry \cite{HK}, which
relates the $\mathcal{N}=2$ version of Liouville theory with the
Kazama-Susuki $SL(2,\mathbb{R})/U(1)$ model. In \cite{M}, Maldacena
explained how the bosonic FZZ\ duality emerges as a consequence of the
supersymmetric extension. More recently, a proof of the FZZ conjectured
duality was given by Hikida and Shomerus in Ref. \cite{HS}; see also \cite%
{GL}.

FZZ duality is a statement about the identity of correlation functions of
both models. It states that correlation functions (\ref{aa})-(\ref{bb}) (or,
equivalently, (\ref{aaa})-(\ref{bbb})) coincide with the sine-Liouville
correlation functions involving the vertex operators%
\begin{equation}
V_{j,m,\bar{m}}(z,\bar{z})=e^{\sqrt{\frac{2}{k-2}}j\varphi (z,%
\bar{z})}e^{i\sqrt{\frac{2}{k}}(mX_{L}(z)+\bar{m}X_{R}(\bar{z}%
))}.  \label{theV}
\end{equation}

The interaction term in sine-Liouville action can actually be written as
particular cases of (\ref{theV}),\ namely%
\begin{equation*}
\lambda \int d^{2}z\ e^{-\sqrt{\frac{k-2}{2}}\varphi }\cos \left( \sqrt{k/2}%
\tilde{X}\right) =\frac{\lambda }{2}\int d^{2}z\ V_{1-\frac{k}{2},\frac{k%
}{2},-\frac{k}{2}}+\frac{\lambda }{2}\int d^{2}z\ V_{1-\frac{k}{2},-\frac{k}{%
2},\frac{k}{2}}.
\end{equation*}

Correlation functions involving (\ref{theV}) were computed in Ref. \cite{FH} using the Coulomb gas approach. The correlators are defined by inserting $s_{+}$ operators of the type $\frac{\lambda }{2}%
\int d^{2}z\ V_{1-k/2,k/2,-k/2}$ and $s_{-}$ operators of the type $\frac{%
\lambda }{2}\int d^{2}z\ V_{1-k/2,-k/2,k/2}$, satisfying the condition
\begin{equation}
\sum\limits_{i=1}^{N}j_{i}+1=\frac{k-2}{2}(s_{+}+s_{-}),\quad \text{with}%
\quad s_{\pm }\in \mathbb{Z}.  \label{18}
\end{equation}

In this theory, the total winding number $\sum_{i=1}^{N}\omega _{i}=\sum_{i=1}^{N}(m_i-\bar{m}_i)/k$ in a given $N$-point function can be violated up to $N-2$ units \cite{FZZ}.
The winding number preserving correlation functions correspond to the particular cases $s_{+}=s_{-}=(\sum_{i=1}^{N}j_{i}+1)/(k-2)$. On the other hand, correlators with $\sum_{i=1}^{N}\omega _{i}\neq 0$ correspond to correlators computed with $%
s_{+}-s_{-}\neq 0$, so that the quantity $(\sum_{i=1}^{N}j_{i}+1)/(k-2)$ in
the latter case is not necessarily an integer
number. In the case of the two-point function ($N=2$) the winding number is preserved, and thus $s_-=s_+$. This implies that one can describe the sine-Liouville 2-point correlation functions by inserting $s_+ + s_-$ operators $e^{-\sqrt{\frac{k-2}{2}}\varphi }\cos ( \sqrt{k/2}%
\tilde{X})$. It can be shown \cite{KKK} that, in this case, the integration over the zero-mode of $\varphi $ in the sine-Liouville two-point function yields a factor\footnote{Here, the symbol $\simeq$ means that this has to be understood as valid in the limit $p_{\phi} >>\sqrt{k-2}$.}
\begin{equation}
\Gamma\left(-2\frac{2j+1}{k-2}\right)
\simeq 
\sqrt{2\pi} e^{-\frac{2i p_{\phi }}{\sqrt{k-2}}( \log p_{\phi } - 1 +\log 2 -\frac{1}{2}\log (k-2))} e^{-\frac{\pi p_\phi}{\sqrt{k-2}}} e^{-\frac{1}{2}\log \left(\frac{2p_\phi}{\sqrt{k-2}}\right)} e^{\frac{i\pi}{4}},  \label{Gamon}
\end{equation} 
and, as observed in \cite{este}, this factor precisely reproduces the leading piece $\sim p_{\phi }\log p_{\phi }$ of the phase-shift\footnote{Although the coefficient of the term in $\delta $ that is linear in $p_{\phi } $ differs from that in (\ref{phaseshift}).} (\ref{phaseshift}). In addition, is was observed there that the integral over the zero mode of $\varphi $ that in the case of sine-Liouville theory produces the $\Gamma $-function (\ref{Gamon}), receives dominant contributions coming from a region of the moduli space that, in the 2D Euclidean black hole $\sigma $-model side, has no representative. This suggests that sine-Liouville description is the approrpiate one to describe these finite $\alpha '$-effects. To complete the argument, in the next section we will address the following two questions: First, whether (and how) finite $\alpha '$-effects (finite-$k$ effects) can be gathered in the 2D Euclidean black hole $\sigma $-model by the integration over $\phi_0 $. Secondly, whether (or to what extent) such $\alpha '$-effects recover the phase-shift (\ref{phaseshift}).

\section{Worldsheet instantons}

The conformal field theory defined by the Kac-Moody currents (\ref{J1})-(\ref%
{J3}) and the stress-tensor (\ref{T}) admits another local exactly marginal
operator. This is given by
\begin{equation}
S_{\text{II}}=\tilde{M}\int d^{2}z\ \tilde{W}(z,\bar{z}), \ \ \ \ \ \tilde{W}={W}^{k-2}=\left( \beta (z) \bar{\beta }(\bar{z})%
\right) ^{k-2}e^{-\sqrt{2(k-2)}\phi (z,\bar{z})}
.  \label{SII}
\end{equation}

Indeed, one can verify that the OPE\ between (\ref{SII}) and currents (\ref%
{J1})-(\ref{J3}) is regular, up to total derivatives. In particular, it
yields%
\begin{equation*}
J^{-}(z)\ \beta ^{k-3}(w)e^{-\sqrt{2(k-2)}\phi (w)}\times h.c.\simeq
\partial _{w}\left( \frac{k-2}{z-w}\beta ^{k-3}(w)e^{-\sqrt{2(k-2)}\phi
(w)}\right) \times h.c.\ +...
\end{equation*}%
where $h.c.$ stands for the Hermitian conjugate contribution. Operator (\ref%
{SII}) commutes with the BRST charge (\ref{BRST}), and has dimension $1$
with respect to the stress-tensor (\ref{T}). Then, one can in principle
consider operators $S_{\text{II}}$ as the screening operator to be used to
define the correlation functions. From the two-dimensional
conformal field theory point of view, the inclusion of
operators (\ref{SII}) is completely natural, as it corresponds to a suitable
screening charge in the Coulomb gas representation of 2D CFT\ correlation
functions. However, within the context of the path integral approach, their
inclusion is less clear as their interpretation in the worldsheet string $
\sigma $-model is neither that of a graviton nor of a tachyon condensate. We will
argue below that operators (\ref{SII})\ are actually
associated to a type of worldsheet instantons discussed in Ref. \cite{MO3}.

If one includes both operators (\ref{SI}) and (\ref{SII})\ in the
correlators, then one finds the
following condition for the observables not to vanish%
\begin{equation}
s+\tilde{s}(k-2)=1+\sum\limits_{i=1}^{N}j_{i},  \label{scompleta}
\end{equation}%
where $s$ and $\tilde{s}$ are the amount of operators of the type $S_{%
\text{I}}$ and of the type $S_{\text{II}}$, respectively.

It was proven in \cite{GN3} that the correlation functions (\ref{A2})
computed with $s$ screening operators of the type (\ref{SI}) and no
operators of the type (\ref{SII}) exactly agree with the correlation
functions computed by inserting $\tilde{s}=s/(k-2)$ operators of the
type (\ref{SII}) and no operators of the type (\ref{SI}) provided the
couplings $M$ and $\tilde{M}$ are related by 
\begin{equation}
\tilde{M}=c_{k}\ M^{k-2},\quad \text{with}\quad c_{k}=\pi ^{k-3}\frac{%
\Gamma \left( k-1\right) }{\Gamma \left( -k+2\right) }\left( \frac{\Gamma
\left( \frac{1}{k-2}\right) }{\Gamma \left( \frac{k-3}{k-2}\right) }\right)
^{k-2}.  \label{M}
\end{equation}

This means that we can write amplitudes (\ref{aa})-(\ref{bb}) as follows%
\begin{equation}
A_{m_{1}m_{2}...m_{N}}^{j_{1}j_{2}...j_{N}}=\Gamma (-\tilde{s})\ c_{k}^{%
\tilde{s}}\ M^{s}\int \prod\limits_{i=4}^{N}d^{2}z_{i}\int
\prod\limits_{r=1}^{\tilde{s}}d^{2}w_{r}\ \tilde{F}%
_{m_{1}m_{2}...m_{N}}^{j_{1}j_{2}...j_{N}}(z_{1},...z_{N};w_{1},...w_{s})
\label{aaa}
\end{equation}%
where
\begin{equation}
\tilde{F}%
_{m_{1}m_{2}...m_{N}}^{j_{1}j_{2}...j_{N}} =
\left\langle \prod\limits_{i=1}^{N}:V_{j_{i},m_{i},\bar{m}_{i}}(z_{i},%
\bar{z}_{i}):\prod\limits_{r=1}^{\tilde{s}}: \tilde{W} (w_r , \bar{w}_{r}):\right\rangle _{M=0}  \label{bbb}
\end{equation}%
and where the amount of screening operators is given by
\begin{equation}
\tilde{s}=\frac{1}{(k-2)}\left( 1+\sum\limits_{i=1}^{N}j_{i}\right) ,
\label{stilde}
\end{equation}%
where the prefactor $\Gamma (-\tilde{s})$ in (\ref{aaa}), which has an
origin completely analogous to the factor $\Gamma (-s)$ in (\ref{aa}), now
exhibits poles at%
\begin{equation}
1+\sum\limits_{i=1}^{N}j_{i}=n(k-2),\qquad n\in \mathbb{Z}_{\geq 0}.
\label{pc}
\end{equation}

Before discussing this prefactor in relation to the phase-shift, let us discuss condition (\ref{pc}) in relation to the worldsheet instantons. Condition (\ref{pc}) for $n=1$, in the case of the $N$-point function,
yields the pole condition%
\begin{equation}
1+\sum\limits_{i=1}^{N}j_{i}=(k-2),
\end{equation}%
which in the convention of \cite{MO3}, which is related to ours through the
simple changes of variables $j_{i}\rightarrow \hat{j}_{i}=j_{i}+1$, and
for $n=1$ correponds to%
\begin{equation}
\sum\limits_{i=1}^{N}\hat{j}_{i}=k+N-3,\qquad n\in \mathbb{Z}_{\geq 0}.
\label{mmm1}
\end{equation}

Interestingly, this is exactly the pole condition that Maldacena and Ooguri
conjectured in \cite{MO3} that would appear in the $SL(2,\mathbb{R})$ WZW\ $N
$-point correlation functions. One confirms from (\ref{mmm1}) that such
poles actually appear. In \cite{MO3}, these poles were interpreted as
worldsheet instantons, corresponding to classical string configurations of momentum $j\sim k$ that
can extend to the large $\phi $ region with no cost of energy. These
configurations correspond to holomorphic maps $\gamma =\gamma (z)$,
associated to classical solutions that extend in the $\phi $ direction with
no potential preventing them from going to $\phi =\infty $. When integrating
over auxiliary fields $\beta $ and $\bar{\beta }$, one produces an
effective potential $\int d^{2}z\partial \bar{\gamma }\bar{%
\partial }\gamma e^{\sqrt{\frac{2}{k-2}}\phi }$, which vanishes for
configurations with $\bar{\partial }\gamma =0$ (as operators (\ref{SI}) and (\ref{SII}) do). These classical
configurations are closely related to the long strings discussed in \cite{SW}.

Poles (\ref{pc}) with $n\geq 1$ also admit an interpretation as worldsheet
instantons. In \cite{MO3}, such divergences are discussed for the particular
case of the two-point function, namely for 
\begin{equation}
\hat{j}=\frac{n}{2}(k-2)+\frac{1}{2},\qquad n\in \mathbb{Z}_{\geq 0},  \label{mmm}
\end{equation}%
and are understood as worldsheet instantons wrapping on $S^{2}$
parameterized by the projective plane variables $\gamma $, $\bar{\gamma 
}$. 

Now, going back to the phase-shift, we observe that for $n=1$ poles (\ref{mmm}) ({\it i.e.} (\ref{mmm1}) for $N=2$) have their origin in the prefactor
\begin{equation}
\Gamma(-\tilde{s}) \simeq 
\sqrt{2\pi} e^{-\frac{i p_{\phi }}{\sqrt{k-2}}( \log p_{\phi } - 1 -\frac{1}{2}\log (k-2))} e^{\frac{i\pi}{4}} e^{-\frac{\pi p_\phi}{2\sqrt{k-2}}} e^{-\frac{1}{2}\log\frac{p_\phi}{\sqrt{k-2}}},  \label{ppp}
\end{equation}
of (\ref{aaa}), which looks pretty much the contribution that yields the stringy phase-shift, cf. (3.3) in \cite{este}. However, a closer look at (\ref{ppp}) reveals that there is a factor $2$ missing with respect to (\ref{phaseshift}). That is, the factor $\Gamma(-\tilde{s})$ that comes from the integration over the dilaton zero-mode once the second screening operator is introduced yields a $k$-dependent function of the momenta that only accounts for {\it one half} of the leading order modification that the phase-shift suffers at high energy. In turn, one concludes, with the authors of \cite{este}, that at high energy the physics of the problem is better described by the FZZ dual model, even if the operator that controls the worldsheet instantons are considered in the 2D Euclidean black hole $\sigma $-model CFT.
\[
\]

Work partially funded by FNRS-Belgium (convention FRFC PDR T.1025.14 and
convention IISN 4.4503.15), by the Communaut\'{e} Fran\c{c}aise de Belgique
through the ARC program and by a donation from the Solvay family. The support of CONICET, FNRS+MINCyT, FONDECyT and UBA through grants PIP 0595/13, BE 13/03, Fondecyt 1140155 and UBACyT 20020120100154BA, respectively, is greatly acknowledged. The Centro de Estudios Cient\'{\i}ficos (CECs) is funded by the Chilean Government through the Centers of Excellence Base Financing Program of CONICYT-Chile.

\end{document}